\documentclass[a4paper]{spie} 
\usepackage[]{graphicx}
\usepackage{amssymb,amsmath, psfrag}
\usepackage{algorithm}
\usepackage{algorithmic}
\usepackage{psfrag} 
\newcommand{\cnum}{\mathbb{C}}

\newcommand{\rnum}{{\bf {R}}}

\newcommand{\curl}{\nabla \times \;}
\newcommand{\curlkz}{\nabla_{k_z} \times \;}
\newcommand{\divo}{\nabla \cdot \;}
\newcommand{\divokz}{\nabla_{k_z} \cdot \;}
\newcommand{\nablakz}{\nabla_{k_z} \;}

\newcommand{\Matrix}[1]{{\mathrm{#1}}}
\newcommand{\Hcurlkz}[1]{H_{k_z}(#1, {\rm curl})}

\newcommand{\Conj}[1]{{\overline{#1}}} 
 
\newcommand{\MyField}[1]{{\bf #1}}

\def\squareforqed{\hbox{\rlap{$\sqcap$}$\sqcup$}}
\def\qed{\ifmmode\else\unskip\quad\fi\squareforqed}

\title{Goal Oriented Adaptive Finite Element Method for the Precise Simulation of Optical Components}
\author{Lin Zschiedrich\supit{a\,b}, Sven Burger\supit{a\,b}, Jan Pomplun\supit{a\,b}, and Frank Schmidt\supit{a\,b}
\skiplinehalf
\supit{a} Zuse Institute Berlin (ZIB), Takustra{\ss}e 7, D-14195 Berlin, Germany \\
\supit{b} JCMwave GmbH, Haarer Stra{\ss}e 14a, D-85640 Putzbrunn, Germany
}
\authorinfo{Further author information: (Send correspondence to Lin Zschiedrich) \\
E-mail: zschiedrich@jcmwave.com \\
URL: http://www.zib.de/nano-optics/ \\ 
URL: http://www.jcmwave.com
}

\begin{document} 
\maketitle 
%%%%%%%%%%%%%%%%%%%%%%%%%%%%%%%%%%%%%%%%%%%%%%%%%%%%%%%%%%%%% 
%% SPIE Copyright form 
\noindent
Copyright 2007  Society of Photo-Optical Instrumentation Engineers.\\
This paper has been published in Proc.~SPIE {\bf 6475}, 6475-16
(2007),  
({\it Integrated Optics: Devices, Materials, and Technologies XI,
Sidorin, Waechter Eds.})
and is made available 
as an electronic reprint with permission of SPIE. 
One print or electronic copy may be made for personal use only. 
Systematic or multiple reproduction, distribution to multiple 
locations via electronic or other means, duplication of any 
material in this paper for a fee or for commercial purposes, 
or modification of the content of the paper are prohibited.
%%%%%%%%%%%%%%%%%%%%%%%%%%%%%%%%%%%%%%%%%%%%%%%%%%%%%%%%%%%%% 

%%%%%%%%%%%%%%%%%%%%%%%%%%%%%%%%%%%%%%%%%%%%%%%%%%%%%%%%%%%%% 
\begin{abstract}
  Adaptive finite elements are the method of choice for accurate simulations of optical components. However as shown recently by Bienstman et al. many finite element mode solvers fail to compute the propagation constant's imaginary part of a leaky waveguide with sufficient accuracy. In this paper we show that with a special {\em goal oriented} error estimator for capturing radiation losses this problem is overcome. 
\end{abstract}
%%%%%%%%%%%%%%%%%%%%%%%%%%%%%%%%%%%%%%%%%%%%%%%%%%%%%%%%%%%%%
\section{INTRODUCTION}
\label{sect:intro}
\begin{figure}
 \psfrag{w}{$W$}
 \psfrag{D}{$D$}
 \psfrag{H}{$H$}
 \psfrag{n1}{$n_{1}=1.0$}
 \psfrag{n2}{$n_{2}=3.5$}
 \psfrag{n3}{$n_{3}=1.45$}
  \begin{center}
      \includegraphics[width=5cm]{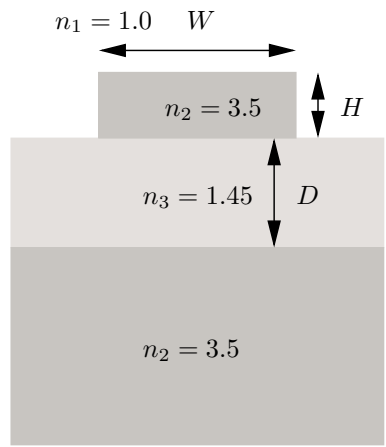}
  \end{center}
  \caption{
\label{Fig:CostWaveguide}
Leaky waveguide as discussed by Bienstman et al.~\cite{Bienstman:06a} The substrate--spacer--air layer stack is considered as infinite. Transparent boundary conditions are needed along the whole boundary.
}
\end{figure}
\begin{figure}
  \begin{center}
      \includegraphics[width=10cm]{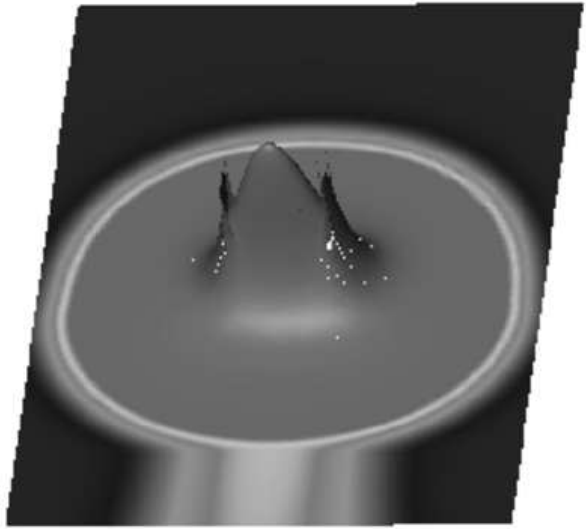}
  \end{center}
  \caption{
\label{Fig:ModeField}
Electric field's absolute value of the fundamental propagation mode for a waveguide as depicted in Figure~\ref{Fig:CostWaveguide} with $H=220 {\mathrm nm}$, $D=1 {\mathrm \mu m}$ and $W=0.5 {\mathrm \mu m}.$ The field is strongly confined in a region near the waveguide core. However, in this pseudocolored plot one perceives a reflection-free (no amplitude modulations) wave propagation in the substrate.   
}
\end{figure}

   
Radiation losses play an important role in the design of open resonator cavities and optical waveguides. Typically the field is strongly confined within such a resonant optical device. A prominent example is a leaky waveguide as discussed by Bienstman et al.~\cite{Bienstman:06a} Figure~\ref{Fig:CostWaveguide} shows the corresponding waveguide geometry. The waveguide core is mounted on a spacer between the substrate and air. The electromagnetic field of the waveguide mode is strongly confined in a region around the waveguide core, Figure~\ref{Fig:ModeField}. Within the air and the spacer the field decays exponentially whereas the field propagates in the substrate. This wave propagation causes the radiation losses. The field amplitude is several orders of magnitude smaller within the substrate block than in the core. Therefore the imaginary part of the propagation constant is much smaller than the real part. For a precise computation of the propagation constant's imaginary part a sufficient accurate simulation of the electromagnetic field within the substrate is needed. Furthermore the transparent boundary condition needs to be adapted. Standard adaptive finite element methods adapt the mesh such that the numerical solutions on successive meshes converge with optimal order in the energy norm. That is, the mesh is adapted such that the error
\begin{eqnarray*}
\delta_{h} & = & \int_{\Omega} \Conj{(\MyField{E}-\MyField{E}_{h})} \epsilon (\MyField{E}-\MyField{E}_{h}) +
 \Conj{(\MyField{H}-\MyField{H}_{h})} \mu (\MyField{H}-\MyField{H}_{h})
\end{eqnarray*}    
is minimized within the computational domain. Here $(\MyField{E}, \MyField{H})$ is the electromagnetic field and $(\MyField{E}_{h}, \MyField{H}_{h})$ its finite element approximation. The real part of the propagation constant shows an optimal convergence as well. Since the imaginary part is small it may happen that the imaginary part is not improved at all when the mesh is refined. This is not a failure of the adaptive mesh refinement strategy because the error estimator was not designed for capturing the radiation losses. Hence one should apply a goal-oriented error estimator as proposed by Rannacher et al.~\cite{Becker:01a} for various applications in engineering. This concept is based on a goal functional $j(\MyField{E})$ (quantity of interest). A special mesh refinement strategy is used to minimize the error
\begin{eqnarray*}
\delta & = & |j(\MyField{E}) - j(\MyField{E}_{h})|.
\end{eqnarray*}  

In the Section~\ref{Sec:NumExp} we will show that a special error estimator for capturing radiation losses allows for a robust and precise simulation of the imaginary part of the propagation constant. Before  we will survey the finite element method and the goal oriented error estimation concept in the sections~\ref{Sec:FEM}--~\ref{Sec:RL}. Functional expressions for the radiation losses which serve as the goal functional are derived in the Sections~\ref{Sec:RWS} and~\ref{Sec:RL}. We also address the construction of goal oriented error estimators for light scattering simulations.
\section{MAXWELL'S EQUATION ON UNBOUNDED DOMAINS AND FINITE ELEMENT DISCRETIZATION}
\label{Sec:FEM}
We start from the time-harmonic Maxwell's equations for the electric field
\begin{eqnarray*}
\curl \mu^{-1} \curl \MyField{E}-\omega^{2} \epsilon \MyField{E} & = & -i \omega \MyField{J} \\
\epsilon \divo \MyField{E} & = & 0. \\
\end{eqnarray*}
Here the electric field $\MyField{E}$ and the material tensors are considered to be defined on the infinite space $\rnum^{3}.$ In this paper we restrict ourselves to waveguide geometries which exhibit a spatial invariance in the $z-$ direction. Furthermore, the electromagnetic field as well the source current \MyField{J} should depend harmonically on $z,$
\begin{eqnarray*}
\MyField{E}(x, y, z) & = & \widetilde{\MyField{E}}(x, y) e^{ik_{z}z} \\ 
\MyField{J}(x, y, z) & = & \widetilde{\MyField{J}}(x, y) e^{ik_{z}z} 
\end{eqnarray*} 
with a propagation constant $k_{z}.$ In the following we drop the wiggles to keep the notation simple. With the definition $\nablakz = (\partial_{x}, \partial_{y}, k_{z})^{T}$ the time-harmonic Maxwell's equations now read as
\begin{subequations}
\begin{eqnarray}
\label{Eqn:MaxwellKZPrim}
\curlkz \mu^{-1} \curlkz \MyField{E}(x, y)-\omega^{2} \epsilon \MyField{E}(x, y) & = &  -i \omega \MyField{J}(x, y) \\
\epsilon \divokz \MyField{E}(x, y) & = & 0. 
\end{eqnarray}
\end{subequations}
\begin{figure}
  \psfrag{PML}{PML}
  \psfrag{rho}{$\rho$}
  \psfrag{eta}{$\eta$}
  \psfrag{xi}{$\xi$}
  \psfrag{omegarho}[2]{}
  \begin{center}
\hspace{3cm}
      \includegraphics[width=6cm]{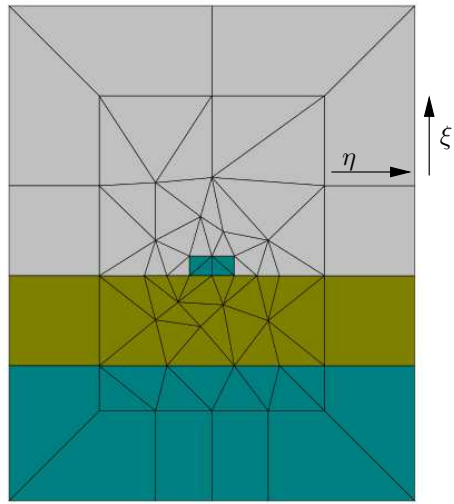}
  \end{center}
  \caption{
Triangulation of the computational domain $\Omega$ with PML. In the exterior domain we introduce a special $(\xi, \eta)-$ coordinate system such that no material jump occurs in $\xi-$direction. The boundaries of the quadrilaterals corresponds to isolines of the $(\xi, \eta)$ coordinate system. The exterior domain is truncated at $\xi=\rho$ whichs yields a bounded domain $\Omega_{\rho}.$
\label{Fig:OmegaRho}
}
\end{figure}   
For the finite element method to be applicable it is necessary to bring Maxwell's equations into a variational form. To this end one needs to define a {\em state space} of admissible fields $\MyField{E}.$  One encounters the difficulty that the fields of interest $\MyField{E}$ have typically infinite field energies on the whole space $\rnum^{3}.$ For example, a propagating mode of a leaky waveguide is exponentially {\em increasing} in certain directions towards infinity.  Furthermore, above Maxwell's equations lack a boundary condition at infinity so far. Hence one could aim to define a {\em state space} which incorporates the outward radiation condition and yields integrable expressions in the variational  formulation of Maxwell's equations. This is indeed done in the Pole condition approach, see \cite{Schmidt2002a,Hohage03a}. However, in this paper we use the perfectly matched layer method (PML). The PML method was originally invented by Berenger \cite{Berenger94} and analyzed and improved by several other authors \cite{Collino:98a,Lassas:01a,Lassas:01b,Hohage03b}. The essential idea is to replace the field $\MyField{E}$ outside a computational domain $\Omega$ by a complex continuation $\MyField{E}_{B}$ chosen such that one observes an exponential decay in the outward direction. The complex continued field $\MyField{E}_{B}$ satisfies Maxwell's equations with modified material tensors $\mu$ and $\epsilon$ outside the computational domain. Due to its exponential decay it is allowed to truncate $\MyField{E}_{B}$ to a finite domain $\Omega_{\rho}.$ Here, $\rho$ is a discretization parameter which corresponds to the thickness of the artificial sponge layer $\Omega_{\rho} \setminus \Omega,$ c.f. Figure~\ref{Fig:OmegaRho}. It can be shown that the truncation error disappears exponentially fast with increasing $\rho$ for certain types of geometries~\cite{Lassas:01a,Lassas:01b,Hohage03b}. To deal with inhomogeneous domains it is necessary to define a special coordinate system $(\eta, \xi)$ in the exterior domain to guarantee the analyticity of $\MyField{E}$ in the outward direction $\xi.$ Especially no material jump is allowed in the $\xi-$ direction, see paper~\cite{Zschiedrich2005a}. To adjust the PML thickness we use an automatic strategy as proposed in our paper~\cite{Zschiedrich:06a}. The variational form is now given as
\begin{eqnarray*}
\label{Eqn:MaxwellVar}
\int_{\Omega_{\rho}} \Conj{\curlkz \MyField{\Phi}} \cdot \widetilde{\mu}^{-1} \curlkz \MyField{E} - \omega^{2}  \Conj{\MyField{\Phi}} \cdot  \widetilde{\epsilon} \MyField{E} & = & -i \omega \int_{\Omega} \Conj{\MyField{\Phi}} \cdot  \MyField{J} \qquad \forall \MyField{\Phi} \in \Hcurlkz{\Omega_\rho}
\end{eqnarray*}
or in short notation
\begin{eqnarray*}
a(\MyField{\Phi}, \MyField{E}) & = & f(\MyField{\Phi}) \qquad \forall \MyField{\Phi} \in \Hcurlkz{\Omega_\rho}
\end{eqnarray*}
with accordingly defined bilinear form $a(\cdot, \cdot):  \Hcurlkz{\Omega_\rho} \times  \Hcurlkz{\Omega_\rho} \rightarrow \cnum$ and functional $f: \Hcurlkz{\Omega_\rho} \rightarrow \cnum$. Here $\Hcurlkz{\Omega_\rho}$ is the space of fields $\MyField{E}(x, y) \in \cnum^{3}$ with finite electric and magnetic field energy within $\Omega_{\rho}.$
The material tensors are modified in the exterior domain to incorporate the PML. The precise definition of the modified material tensors $\widetilde{\mu}$ and $\widetilde{\epsilon}$ in the exterior domain $\Omega_{\rho} \setminus \Omega$ are given in our paper~\cite{Zschiedrich2005a}.
We again drop the wiggles for the sake of a simpler notation. 
In the following we will frequently make use of the formula
\begin{eqnarray}
\int_{\Omega}  \Conj{\curlkz \MyField{E}} \cdot \widetilde{\mu}^{-1} \curlkz \MyField{E} - \omega^{2}  \Conj{\MyField{E}} \cdot  \widetilde{\epsilon} \MyField{E} + i \omega \Conj{\MyField{\Phi}} \cdot  \MyField{J} & = & - 2 \Im \left( k_{z} \right)\int_{\Omega} \left[  \Conj{\MyField{E}} \times \mu^{-1}  \curlkz \MyField{E} \right]\cdot \hat{n}_{z} \nonumber  \\
{} & {} & + \int_{\partial \Omega}  \left[  \Conj{\MyField{E}} \times \mu^{-1}  \curlkz \MyField{E} \right]\cdot \hat{n}_{xy} \label{Eqn:PartIntegration}.
\end{eqnarray}
Here $\hat{n}_{z} = (0, 0, 1)^{T}$ and $\hat{n}_{xy}$ is the normal on $\partial \Omega$ perpendicular to the $z-$ direction. Formula~\eqref{Eqn:PartIntegration} may be derived from Maxwell's equations by a multiplication of equation~\eqref{Eqn:MaxwellKZPrim} with $\Conj{\MyField{E}(x, y)}$ and integration over the three dimensional domain $\Omega \times [0, z_{\delta}]$ followed by partial integration based on Green's formula and afterwards differentiation with respect to $z_{\delta}.$ 

The variational form of Maxwell's equation~\eqref{Eqn:MaxwellVar} is discretized with high-order vector elements $\MyField{\Psi}_h \in V_h \subset \Hcurlkz{\Omega_\rho}.$ Here $V_h$ is a finite dimensional subspace of $\Hcurlkz{\Omega_\rho}$ spanned by the finite element ansatz functions ${\MyField{\Psi}_{1}, \cdot, \MyField{\Psi}_{N}}.$ This gives the discretized variational problem for $\MyField{E}_h \in V_h,$
\begin{eqnarray*}
a(\MyField{\Phi}_h, \MyField{E}_h) & = & f(\MyField{\Phi}_h) \qquad \forall \MyField{\Phi}_h \in V_h, 
\end{eqnarray*}
which can be casted into a linear algebraic system
\begin{eqnarray}
\label{Eqn:MaxAlg}
\left( \Matrix{A}_{0}+k_{z}\Matrix{A}_{1}+k_{z}^{2}\Matrix{A}_{2} - \omega^{2} \Matrix{B} \right) \Matrix{e} & = & \Matrix{f}. 
\end{eqnarray}
Here $\Matrix{e}$ is the coefficient vector $e=(e_{1}, \dots, e_{N})$ of the finite element solution, 
\begin{eqnarray*}
\MyField{E}_{h} & = & \sum_{l=1}^{N}e_{l}\MyField{\Psi}_{l}.
\end{eqnarray*}

\section{RESONANCE MODES, WAVEGUIDE MODES AND SCATTERING SOLUTIONS}
\label{Sec:RWS}
Maxwell's equations give rise to three typical simulation scenarios which we shortly review below. In each case one computes certain quantities of interest from the simulated field $\MyField{E},$ e.g. the electric field energy $E_{\Omega}(\MyField{E})$ within $\Omega,$ the power flux $P_{\Omega}(\MyField{E})$ through the cross section computational domain $\Omega$ and the in plane power flux $P_{\partial \Omega}(\MyField{E})$ across the boundary of the cross section computational domain:
\begin{eqnarray*}
E_{\Omega}(\MyField{E}) & = & \frac{1}{4} \Re \left( \int_{\Omega} \MyField{E} \cdot \Conj{\epsilon \MyField{E}} \right) \\
P_{\Omega}(\MyField{E}) & = & \frac{1}{2\Re (\omega)} \Im\left( \int_{\Omega} \left[  \Conj{\MyField{E}} \times \mu^{-1}  \curlkz \MyField{E} \right]\cdot \hat{n}_{z} \right) \\
P_{\partial \Omega}(\MyField{E}) & = & \frac{1}{2 \Re(\omega)} \Im\left(\int_{\partial \Omega}  \left[  \Conj{\MyField{E}} \times \mu^{-1}  \curlkz \MyField{E} \right]\cdot \hat{n}_{xy} \right) 
\end{eqnarray*}
\subsubsection*{Resonance Modes} 
In this setting one seeks eigenpairs $(\MyField{E}, \omega)$ for given propagation constant $k_{z}$ and $\MyField{J} = 0.$ Numerically this corresponds to an eigenvalue problem for $(\Matrix{e}, \omega^2)$ in the matrix equation~\eqref{Eqn:MaxAlg}. We want to relate the imaginary part of $\omega$ to the functionals above. Assuming transparent materials in the computational domain $\Omega$ and $\Im({k_{z}})=0$ one derives from formula~\eqref{Eqn:PartIntegration} that 
\begin{eqnarray}
\label{Eqn:ImOmega}
\Im \left( \omega \right) = -2\frac{P_{\partial \Omega}(\MyField{E})}{E_{\Omega}(\MyField{E})}.
\end{eqnarray}
Hence up to normalization the imaginary part of the resonance mode $\omega$ is proportional to the in-plane losses of the resonator. Therefore construction of a goal orientated error estimator for the computation of the qualtity factor of a cavity should start from~\eqref{Eqn:ImOmega}.  
\subsubsection*{Waveguide Modes}
In this setting one seeks pairs  $(\MyField{E}, k_{z})$ for a prescribed angular frequency $\omega \in \rnum$ such that equation~\eqref{Eqn:MaxwellVar} is satisfied with $\MyField{J} = 0.$ Now as follows from equation~\eqref{Eqn:PartIntegration} the imaginary part of $k_{z}$ is given by
\begin{eqnarray}
\label{Eqn:ImKZ}
\Im \left( k_{z} \right) = \frac{P_{\partial \Omega}(\MyField{E})}{2 P_{\Omega}(\MyField{E})}. 
\end{eqnarray}
So the imaginary part of $k_{z}$ is exactly given as the ratio of the in plane power flux across the boundary of the computational domain and the power flux within the computational domain in waveguide direction.
\subsubsection*{Scattering Solutions} 
In this setting $\omega$ as well as $k_{z}$ are prescribed and either the source term $\MyField{J}$ is non-zero or an incident field is present. It is described in~\cite{Zschiedrich:06a} how to incorporate an incident field into the variational form of Maxwell's equation. Here we restrict ourselves to the special case that the current source consists of a ``line'' dipole source 
\begin{eqnarray*}
\MyField{J}(x, y, z) = \delta(x-x_0, y-y_0)e^{ik_z z}\MyField{J}_{0}
\end{eqnarray*} 
with $(x_{0}, y_{0})$ in $\Omega.$ In this case the regularity of the solution is poor and the finite element method will suffer from a slow convergence. To cure that we use the subtraction approach, where the total field is split into an analytically given singular field $\MyField{E}_{s}$ and a correction field $\MyField{E}_{c}$ with smoothness properties fitting well in the finite element concept. We assume that the line source is not placed on a material jump, so that $\mu_{0} = \mu(x_{0}, y_{0})$ and $\epsilon_{0} = \epsilon(x_{0}, y_{0})$ are well defined. The solution $\MyField{E}_{s}$ to Maxwell's equation 
\begin{eqnarray*}
\curlkz \mu^{-1}_{0} \curlkz \MyField{E_s}(x, y)-\omega^{2} \epsilon_{0} \MyField{E_s}(x, y) & = &  -i \omega  \delta(x-x_0, y-y_0)\MyField{J}_{0}
\end{eqnarray*}
with constant material tensors $\mu_{0}$ and $\epsilon_{0}$ is known analytically \cite{Paulus:01a}. Now the correction field satisfies Maxwell's equation with the right hand side term
\begin{eqnarray*}
f_{c} & = &  -\curlkz \left(\mu^{-1} - \mu^{-1}_{0} \right) \curlkz \MyField{E_s}(x, y)+\omega^{2} \left(\epsilon-\epsilon_{0} \right) \MyField{E_s}.
\end{eqnarray*}
Since $\mu^{-1} - \mu^{-1}_{0}$ and $\epsilon-\epsilon_{0}$ are zero at $(x_{0}, y_{0})$ it can be proven that the right hand side $f_{c}$ is sufficiently smooth for an optimal convergence rate of the finite element method. The right hand side $f_{c}$ might be non-zero everywhere especially in the exterior domain. Due to the analyticity of $f_{c}$ the additional source terms will cause no problem in the PML. Since many finite element implementations do not allow for a source term in the PML one alternatively may compute the total field in the exterior domain. In this case additional boundary terms on $\partial \Omega$ stemming from the singular field are needed. This is similar to the treatment of coupling with an incoming field and is naturally implemented within the finite element method.

\section{GOAL ORIENTED ERROR ESTIMATOR: GENERAL CONCEPT}
We follow the general concept for the construction of goal oriented error estimators as proposed by Rannacher~\cite{Becker:01a}. The quantity we aim to compute is a functional $j(\MyField{E})$ of the solution field $\MyField{E}.$ So the finite element mesh should be adapted such that the error $j(\MyField{E})-j(\MyField{E}_{h})$ is efficiently reduced. The idea by Rannacher is to embed this situation into the framework of optimal control theory. To this end, we define the {\em trivial} constrained optimization problem
\begin{eqnarray}
j(\MyField{E})-j(\MyField{E}_{h}) = \min_{\MyField{\Psi} \in\ \Hcurlkz{\Omega_\rho}} \left\{j(\MyField{\Psi})-j(\MyField{E}_{h}) \; : \; \; a\left(\MyField{\Phi}, \MyField{\Psi} \right) = f(\MyField{\Phi}) \;  \forall \MyField{\Phi} \in \Hcurlkz{\Omega_\rho} \right\}.
\label{Eqn:MinProblem}
\end{eqnarray}
The Minimum of~\eqref{Eqn:MinProblem} corresponds to a stationary point $(\MyField{E}, \MyField{E^{*}})$ of the Lagrangian 
\begin{eqnarray*}
\mathcal{L}(\MyField{E}, \MyField{E^{*}}) = j(\MyField{E})-j(\MyField{E}_{h}) + f(\MyField{E}^{*})- a\left(\MyField{\MyField{E}^{*}}, \MyField{E} \right)
\end{eqnarray*}          
where $\MyField{E}^{*}$ denotes the 'dual' variable (Lagrangian multiplier). Hence, we seek the solution $(\MyField{E}, \MyField{E}^{*})$ to the Euler-Lagrange system
\begin{subequations}
\label{Eqn:EulerLagrange}
\begin{eqnarray}
a\left(\MyField{\Phi}, \MyField{E} \right) & = & f(\MyField{\Phi})  \qquad \forall \MyField{\Phi} \in \Hcurlkz{\Omega_\rho} \label{Eqn:EulerLagrangePrimary} \\
a\left(\MyField{E}^{*}, \MyField{\Phi} \right) & = & j'(\MyField{E};    \MyField{\Phi})  \qquad \forall \MyField{\Phi} \in \Hcurlkz{\Omega_\rho}. \label{Eqn:EulerLagrangeDual}
\end{eqnarray}
\end{subequations}
The first equation~\eqref{Eqn:EulerLagrangePrimary} is the variational form of original Maxwell's equations. In the dual equation~\eqref{Eqn:EulerLagrangeDual} the functional $j(\MyField{E})$ -- the quantity of interest -- enters the stage in form of its linearization $j'(\MyField{E}).$  A finite element discretization of the Euler-Lagrange system yields a supplemental discrete problem
\begin{eqnarray*}
a\left(\MyField{E}_{h}^{*}, \MyField{\Phi}_h \right) & = & j'(\MyField{E}_{h};    \MyField{\Phi}_{h})  \quad \forall \MyField{\Phi}_h   \in V_{h}. 
\end{eqnarray*}
Introducing the primal and dual residuums, 
\begin{eqnarray*}
\rho(\MyField{E}_{h}; \cdot) & =  & f(\cdot)-a\left(\cdot, \MyField{E} \right) \\
\rho(\MyField{E}^{*}_{h}; \cdot) & =  & j'(\MyField{E}_{h}; \cdot)-a\left(\MyField{E}^{*}_{h}, \cdot \right)
\end{eqnarray*}
one obtains the error representation
\begin{eqnarray}
\label{Eqn:ErrRepr}
 j(\MyField{E})-j(\MyField{E}_{h}) = 
\frac{1}{2}\rho(\MyField{E}_{h}; \MyField{E}^{*}-w_{h})
+\frac{1}{2}\rho^{*}(\MyField{E}^{*}_{h}; \MyField{E}-v_{h})+R
\end{eqnarray}
with remainder term $R$ of third order in $\MyField{E}-\MyField{E}_{h}$ and  $\MyField{E}^{*}-\MyField{E}_{h}^{*}$, see~\cite{Becker:01a}.

Eigenvalue problems of the form 
\begin{eqnarray*}
\label{Eqn:VarEig}
a(\MyField{\Phi}, \MyField{E}) - \sigma b(\MyField{\Phi}, \MyField{E}) & = & 0
\end{eqnarray*}
can also be treated within this approach~\cite{Heuveline2001a}. To do this one extends the state space by the vector space of complex numbers $\cnum$,  $V \rightarrow V \times \cnum.$ Since a solution of an eigenvalue problem is only fixed uniquely up to normalization one includes a normalization, e.g. $b(\MyField{E}, \MyField{E}) = 1$ into the variational form~\eqref{Eqn:VarEig}, 
\begin{eqnarray*}
\label{Eqn:VarEigNormalized}
a(\MyField{\Phi}, \MyField{E}) - \sigma b(\MyField{\Phi}, \MyField{E}) + \Conj{\chi} \left(b(\MyField{E}, \MyField{E})-1 \right) & = & 0. \qquad \forall (\MyField{\Phi}, \chi)   \in \Hcurlkz{\Omega_\rho} \times \cnum.
\end{eqnarray*}
which gives the non-linear variational problem
\begin{eqnarray*}
\widetilde{a}\left( (\MyField{\Phi}, \chi), (\MyField{E}, \sigma) \right) & = & 0.
\end{eqnarray*}
The corresponding dual problem is derived in~\cite{Heuveline2001a}. For a chosen goal functional $j(\MyField{E}), $ it is necessary to check the unique solvability of the dual problem and its discretization by finite elements. For a wide range of goal functionals an error bound of the form
\begin{eqnarray}
\label{Eqn:ErrReprProduct}
|j(\MyField{E}-\MyField{E_h}| \leq C \sum_{T} \|\rho\|_T \|\rho^{*}\|_T
\end{eqnarray}  
can be derived from the error representation~\eqref{Eqn:ErrRepr}, see~\cite{Becker:01a}. Here $\|\rho\|_T$ denotes the norm of the residuum restricted to the triangle $T$. As usual for an adaptive strategy we refine triangles with large local contributions $\|\rho\|_T \|\rho^{*}\|_T$ to the sum, see~\cite{Heuveline2001a}.
\section{GOAL ORIENTED ERROR ESTIMATOR: ADAPTIVE COMPUTATION OF RADIATION LOSSES}
\label{Sec:RL}
In this section we derive a goal oriented error estimator for the precise computation of radiation losses in optical devices. Here we restrict ourselves to waveguide mode problems. So the non-linear functional~\eqref{Eqn:ImKZ} is a natural candidate for the goal functional. But since the nominator on the left hand side of equation~\eqref{Eqn:ImKZ} involves derivatives of the solution field $\MyField{E}$ this functional is not defined on the whole space $\Hcurlkz{\Omega_{\rho}}.$ A regularization of the functional is required. This is done in our case by an appropriate extension of the boundary integral expression into a volume integral around the boundary $\partial \Omega,$  
\begin{eqnarray}
\label{Eqn:WGGoalFunctionalCorrect}
j(\MyField{\Phi}) & = & \frac{\Im \left(\int_{\Gamma}  \left[  \Conj{\MyField{\Phi}} \times \mu^{-1}  \curlkz \MyField{\Phi} \right]\cdot \hat{n}_{x} +   \left[ \Conj{\MyField{\Phi}} \times \mu^{-1}  \curlkz \MyField{\Phi} \right]\cdot \hat{n}_{y} \right)} {\Im\left( \int_{\Omega} \left[  \Conj{\MyField{\Phi}} \times \mu^{-1}  \curlkz \MyField{\Phi} \right]\cdot \hat{n}_{z} \right)}. 
\end{eqnarray}   
In our implementation we simplify this expression by replacing $\curlkz \MyField{\Phi}$ by $\curlkz \MyField{E_h}.$ The so modified functional is unchanged when applied to the solution field $\MyField{E}.$  
\section{A WAVEGUIDE EXAMPLE}
\label{Sec:NumExp}
\begin{figure}
  \psfrag{kz}{$k_{z}$}
  \psfrag{ppppppppppppppppppppPQ}[lc][lc][1.2][0]{$\frac{P_{\partial \Omega}(\MyField{E})}{2 k_{0}P_{\Omega}(\MyField{E})}$}
  \begin{center}
    \includegraphics[width=8cm]{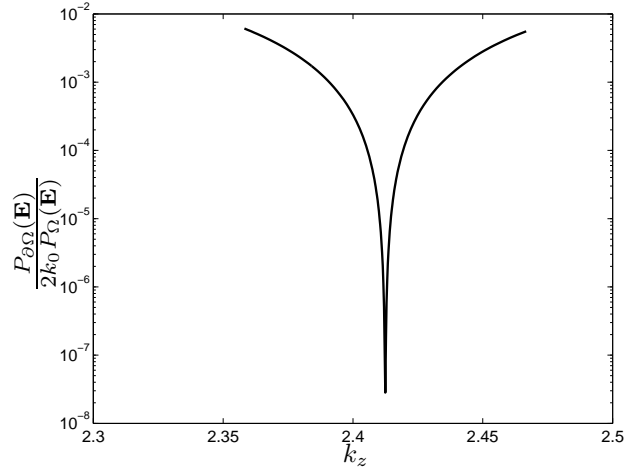} 
  \end{center}
  \caption{
Ratio of the in plane power flux across the boundary of the computational domain and the power flux within the computational domain in waveguide direction. Time harmonic Maxwell's equations are solved for a vacuum wavelength $\lambda_{0} = 1.55 \mu {\mathrm{m}}$ and various $k_{z}.$ As excitation a dipole is placed in the center of the waveguide core. One observe a dramatic reduction of in-plane losses when $k_{z}$ reaches the real part of the propagation mode.  
\label{Fig:KZScan}
}
\end{figure}

We apply our finite element solver to the challenging example of a leaky waveguide as depicted in Figure~\ref{Fig:CostWaveguide}. This example was discussed on the OWTNM 2006 by Bienstman et al.~\cite{Bienstman:06a} and it was shown that an accurate computation of the losses is challenging. We computed the propagating mode with three mesh refinement strategies; uniform mesh refinement, adaptive mesh refinement with energy norm error estimator and adaptive mesh refinement with the new loss functional based error estimator. Third order finite elements were used. The computed eigenmodes are given in the Tables~\ref{Tab:Uniform}--~\ref{Tab:AdaptGoal}. In the first column the refinement steps are given, in the second the number of used unknowns and in the third column the computed effective refractive index. 
The reference value for the effective index of the waveguide is taken from~\cite{Bienstman:06a}, 
\begin{equation}
n_{\mathrm{eff}} = k_{z}/k_{0} = 2.4123720+2.91348 \cdot 10^{-8}i.
\end{equation}
To start the eigenmode solver an initial guess for the propagation constant is needed. Such a guess can be obtained by placing a dipole source in the center of the waveguide for various $k_{z}$, see Figure~\ref{Fig:KZScan}. Time-Harmonic Maxwell's equations are solved and the right hand side of Formula~\eqref{Eqn:ImKZ} is computed. In the resonant regime we observe a dramatic reduction of in-plane losses. The cusp of the curve in Figure~\ref{Fig:KZScan} is an excellent initial guess for the eigenmode solver.     

In all three cases one observes convergence to the reference value. The benefits of an adaptive mesh refinement strategy are significant. As expected with the loss based mesh adaption (Table~\ref{Tab:AdaptGoal}) the imaginary part of the propagation constant converges faster than with the energy based mesh adaption, especially for lower accuracy demands. As can be seen from Table~\ref{Tab:AdaptEnergy} the imaginary part is not improved in the first three refinement steps when the energy based error estimator is used. However, for high accuracy demands they asymptotically show nearby the same behavior. For the loss based adaptive refinement method it takes less than $10$ seconds to reach a relative accuracy of four digits in the imaginary part on a standard PC.   
\begin{table}
\begin{center}
\begin{tabular}{rrrr}
Step & $N^o$ DOF & $\Re(n_{\rm eff})$ &  $\Im(n_{\rm eff})$ \\
\hline 
0 & 3075 & 2.40362444e+00 & 1.76283e-08 \\
1 & 6900 & 2.41244790e+00 & 2.95880e-08 \\
2 & 16530 & 2.41258532e+00 & 2.90962e-08 \\
3 & 43710 & 2.41245837e+00 & 2.91179e-08 \\
4 & 129750 & 2.41240459e+00 & 2.91284e-08 \\
5 & 428550 & 2.41238429e+00 & 2.91324e-08 \\
6 & 1533030 & 2.41237665e+00 & 2.91339e-08
\end{tabular}
\end{center}
\caption{
Computed effective refractive index: uniform mesh refinement. 
\label{Tab:Uniform}
}
\end{table}

\begin{table}
\begin{center}
\begin{tabular}{rrrr}
Step & $N^o$ DOF & $\Re(n_{\rm eff})$ &  $\Im(n_{\rm eff})$ \\
\hline
0 &  3075 & 2.40362444e+00 & 1.76283e-08 \\
1 &  3531 & 2.41244233e+00 & 1.57048e-08 \\
2 &  4581 & 2.41257875e+00 & 1.53788e-08 \\
3 &  7080 & 2.41246067e+00 & 1.59441e-08 \\
4 &  12780 & 2.41240706e+00 & 2.96001e-08 \\
5 &  21918 & 2.41238542e+00 & 2.96067e-08 \\
6 &  40875 & 2.41237710e+00 & 2.91368e-08 \\
7 &  73977 & 2.41237393e+00 & 2.91370e-08 \\
8 &  140631 & 2.41237272e+00 & 2.91348e-08 \\
9 &  253395 & 2.41237226e+00 & 2.91348e-08 \\
10 &  476511 & 2.41237209e+00 & 2.91348e-08 
\end{tabular}
\end{center}
\caption{
\label{Tab:AdaptEnergy}
Computed effective refractive index: energy based mesh refinement. In the first refinement steps the imaginary part is not improved.
}
\end{table}

\begin{table}
\begin{center}
\begin{tabular}{rrrr}
Step & $N^o$ DOF & $\Re(n_{\rm eff})$ &  $\Im(n_{\rm eff})$ \\
\hline
0 &  3075 & 2.40362444e+00 & 1.76283e-08 \\
1 &  4698 & 2.40636085e+00 & 3.10891e-08 \\
2 &  8895 & 2.41259520e+00 & 2.90980e-08 \\
3 &  14718 & 2.41250075e+00 & 2.91101e-08 \\
4 &  23019 & 2.41242571e+00 & 2.91240e-08 \\
5 &  39903 & 2.41239249e+00 & 2.91307e-08 \\
6 &  62856 & 2.41237984e+00 & 2.91333e-08 \\
7 &  112818 & 2.41237497e+00 & 2.91343e-08 \\
8 &  203637 & 2.41237312e+00 & 2.91346e-08 \\
9 &  385551 & 2.41237241e+00 & 2.91348e-08
\end{tabular}
\end{center}
\caption{
Computed effective refractive index: loss based mesh refinement. Imaginary part is improved from the beginning of the mesh refinement process. 
\label{Tab:AdaptGoal}
}
\end{table}

\bibliography{lit1,lit2,other}   
\bibliographystyle{spiebib}  
\end{document}